# Low Energy, High Flux, Uniform and Large Field Size Electron Beam Facility

[1]Ali Behcet Alpat, [2]Giovanni Bartolini, Talifujiang Wusimanjiang, [3]Goesta Mattausch, Tobias Teichmann, Ralf Bluethner,  [4]Michael Thomas Müller, Carsten Zschech, [5]Abdullah Coban, Arca Bozkurt

*Abstract*— **Materials to be deployed in space applications have to undergo a variety of different test scenarios, simulating actual space conditions. Among these materials solar photovoltaic cells, optics, meta-materials and more will be directly exposed to space radiation and must be tested accordingly. From the design phase of such target materials to the final production, it is important to obtain information about their behavior and performance in defined irradiation scenarios and qualify them following the directions of relevant ECSS/ESA standards. An excellent method to cover part of these tests in a laboratory scale is the high-flux, high-fluence electron irradiation with the help of industrial electron beam generators.**

**The present collaboration has developed different irradiation test setups and procedures, adapted to the existing electron beam generators at IPF in Dresden, Germany covering an energy range between 100 keV to 1.5 MeV. The test setups, aiming for high electron flux, uniform and large irradiation field sizes, and their capabilities for use in irradiation qualification tests for space applications are described in this paper.**

## I. INTRODUCTION

The electron irradiation of solar cells, optical coatings and polymer-based metamaterials is of paramount importance to qualify them during all phases from design to production and, subsequently, to check their performance according to ECSS/ESA standards. Recent publications also show the potential of coatings on solar cells to improve their power conversion efficiencies [1]. Next-generation, flexible polymer-based metamaterials are being developed to be used in space applications as for instance, optical solar reflectors [2] that need to be qualified under low energy electron beam irradiation up to high fluences of about $10^{16-17}$ e/cm$^2$. In Europe, there are very few facilities capable of fulfilling the restrictive test requirements. Hence, our work has focused to create two alternative flexible irradiation setups to offer to the community where we have full experimental control within the machine's limits to answer requests, that may arise from the users.

## II. IPF IRRADIATION FACILITY

The Leibniz-Institute of Polymer Research in Dresden, Germany, IPF [3] runs two different electron gun systems located at the same irradiation facility.

The first accelerator is an industrial machine of the series ELV-2 of the Budker Institute of Nuclear Physics, Novosibirsk, Russia (BINP), installed between 1994 and 1996 in Dresden. It operates in the energy range from 0.6 MeV to 1.5 MeV. The beam power maximum is 6 kW and the beam current maximum is 4 mA. The machine uses a fast-scanning electron beam capable of delivering uniform electron flux over a wide area. Dose rates typically range from 4.5 kGy/min to 50 kGy/min (1 MeV in PMMA) corresponding to $2\times10^{11}$ to $5\times10^{12}$ e/(cm$^2$s). Thus, fluences up to $10^{17}$ e/cm$^2$ or higher can be achieved within reasonable irradiation times. The samples can either be in a static position under the beam or, for larger area irradiations, can be placed on a moving table/conveyor system. Table feed and oscillating motions can be programmed to deliver the desired dose distribution.

The second electron accelerator is a low-energy linear emitter, EBE 300/270 (COMET AG). The device is mounted to a programmable industrial robot. In this setup, the samples are placed in a fixed position below the source. The robot arm then is moved over the surface to achieve the wanted dose distribution. Also, the irradiation of 3D-shaped products is possible [4]. Dose rates typically range from 12.5 kGy/min to 150 kGy/min (200 keV electrons in PMMA) or flux of $5\times10^{11}$ to $7\times10^{12}$ e/(cm$^2$s). Fluences up to $10^{17}$ e/cm$^2$ or higher can be achieved. The technical specifications of both electron sources are summarized in Table 1. The machines are shown in Figure 1.

[1]A.B. Alpat is with Istituto Nazionale di Fisica Nucleare, Sezione di Perugia, Via A.Pascoli snc, 06123, Perugia, Italy (e-mail: behcet.alpat@pg.infn.it).
[2]G. Bartolini and T. Osmanjan are with BEAMIDE s.r.l., Via Campo di Marte 4/o, 06124, Perugia, Italy (e-mails: giovanni.bartolini@beamide.com and talipjan.osman@beamide.com), [3]G. Mattausch, T. Teichmann and R. Bluethner are with Fraunhofer FEP, Institute for Organic Electronics, Electron Beam and Plasma Technology, Winterbergstrasse 28, 01277 Dresden, Germany (e-mails: goesta.mattausch@fep.fraunhofer.de, tobias.teichmann@fep.fraunhofer.de, ralf.bluethner@fep.fraunhofer.de),  [4]M.T. Müller and C. Zschech are with Leibniz-Institute of Polymer Research, Hohe Strasse 6, 01069 Dresden, Germany (e-mails: mueller-michael@ipfdd.de, zschech@ipfdd.de), [5]A. Coban, A. Bozkurt and D. Cegil are from IRADETS, A.S. Teknopark İstanbul, Pendik/İstanbul, Turkey (e-mails: abdullah.coban@iradets.com, arca.bozkurt@iradets.com, dogukan.cegil@iradets.com)

| PARAMETERS | EBE-300/270 | ELV-2 |
|---|---|---|
| Accelerating voltage | 100 keV to 300 keV | 0.6 MeV to 1.5 MeV |
| Max. power at max. high voltage | 4.5 kW | 6 kW |
| Surface dose uniformity | < ±10% at 100 keV | < ±10% at 1 MeV |
| Beam width | 230 mm | 800 mm |

Table 1. Technical specifications of EBE-300/270 and ELV-2 at Leibniz-Institute of Polymer Research in Dresden

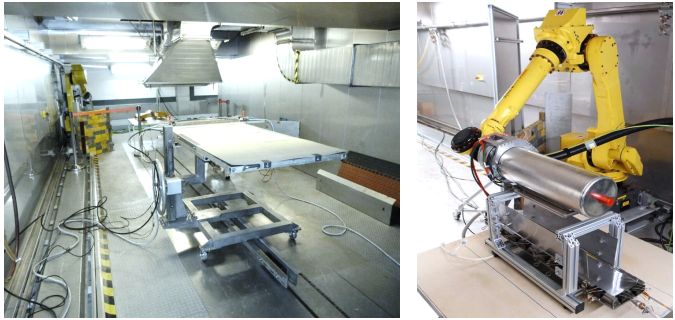

Fig. 1 Irradiation room with electron exit window of ELV-2 over the movable table and robot setup in the background (left). Close-up of the robotic arm and irradiation setup (right). Both sources are located in the same irradiation room.

### III. SETUPS FOR SPACE APLLICATION TESTS

In collaboration with Fraunhofer FEP [5] and BEAMIDE [6] both electron sources have been equipped with setups, optimized to follow the requirements for the test of materials for space applications. These setups have been tested and characterized by extensive dosimetry using radiochromic film (crosslinking AB) on-site as well as with the help of radiation transport simulations (Geant 4 & Fluka).

Two identical custom irradiation boxes of 80 × 10 cm² useful area have been designed, that can be combined with both electron beam setups. The development was focused on the proper mounting, cooling and inertization of the samples during irradiation. This resulted in the following main features:

- Water cooling system for contact cooling of the substrate carrier
- Micro-hole vacuum system to hold the samples in place on the cooling plate
- Nitrogen purging of the sealed irradiation volume (Mylar foil-covered frame with flat seals closes the box to the top and prevents oxygen contamination)
- Contamination monitor for oxygen concentration
- Multiple thermal probes integrated the cooling plate and attached to dummy samples to monitor the sample temperature in real time

Figure 2 shows example monitor data for one 33-minute irradiation. The monitor graph includes warm up procedure of the beam and purging of the irradiation box with nitrogen. When the box is brought under the beam at ca, minute 16, an increase in both center end edge temperature is measured. The oxygen concentration remains constant at low levels over the whole exposure time. The irradiation stops at minute 49.

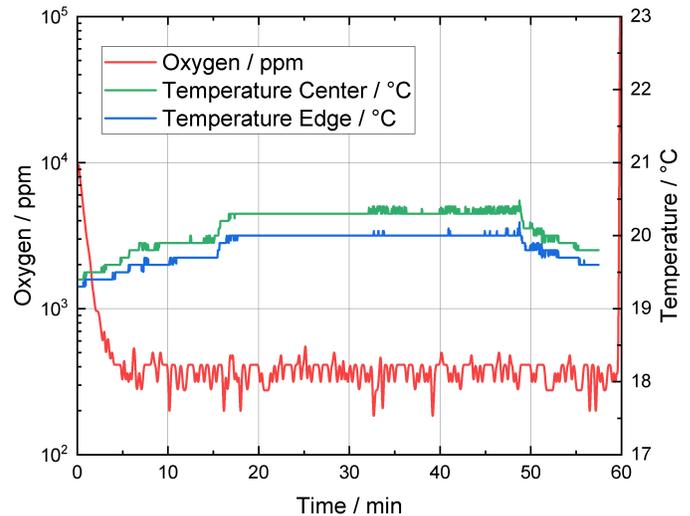

Fig. 2 Example of monitor data for a 33-minute irradiation run at ELV-2

In addition to the moving table, which is also adjustable in height, the ELV-2 can be equipped with a custom slit aperture to provide further attenuation of the electron beam intensity and achieve low flux while maintaining the desired dose homogeneity over a given area. This can be achieved by an oscillating motion of the irradiation box below the aperture setup. The setup is displayed in Figure 3.

To optimize the irradiation parameters: aperture width, oscillation amplitude and beam current, the setup including the electron exit of the ELV-2 has been simulated with Geant4. An impression of the simulation geometry is given in Figure 4.

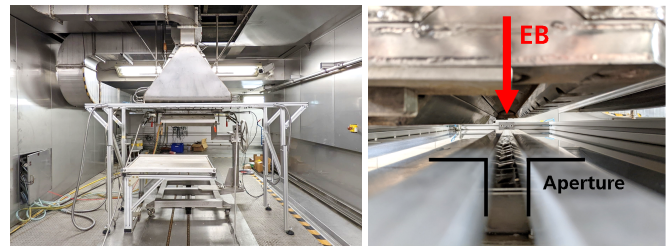

Fig. 3 Aperture slit setup with oscillating irradiation box below ELV-2 (left). Close-up of electron exit window with an adjustable aperture below (right).

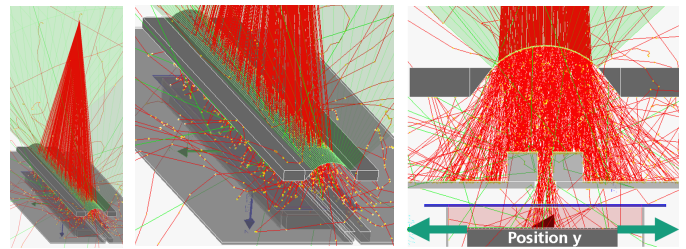

Fig. 4 Geant4 simulation of the ELV-2 scanned beam fan on electron exit window with aperture setup below (left and middle). Close-up side view of the electron exit from vacuum to atmosphere, attenuation through slit aperture and irradiation of inert box with samples (right).

Figure 5 left illustrates the dependency of the dose distribution at the sample surface on the oscillating amplitude for a fixed aperture width. The best configuration has then been used in the setup at the site. The measured dosimetric data are in good agreement with the forecast distribution (Figure 5 right).

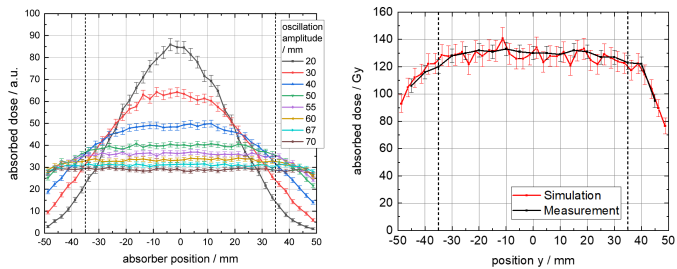

Fig. 5 Results of the Geant4 simulation of the ELV-2. Dose profiles for different oscillating amplitudes of the moving table (left). Comparison of simulation and measured data for the final configuration (right).

For this example, the goal was to achieve a homogenous dose distribution over the full sample width of 70 mm combined with an energy spread below 7.5 % at the sample surface. The setup can be adapted to irradiate much larger flat samples as well while keeping good dose uniformity if the customer agrees to a variation in dose rate during the irradiation.

The custom irradiation box can also be used with the 300 kV electron gun. To that configuration, a reflector setup (Figure 6 left) has been added to improve the dose homogeneity. The influence of dimensions and distances of the reflector surfaces on dose and energy uniformity as well as on the secondary photon creation have been studied in detail for various electron energies. The simulated geometry is shown in Figure 6 right.

The results have been validated by experimental data (agreement within 5%) using radiochromic film placed over the irradiation area.

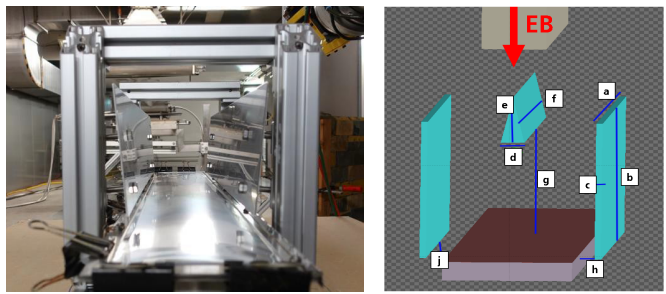

Fig. 6 The electrons exit window (left), side reflectors and bottom irradiation box and its simulation with Geant4 (right).

The described tool set allows to create user-specific running conditions (energies, flux and fluences) and to perform tests compliant with relevant ECSS/ESA standards.

To obtain an electron energy of 200 keV at the target surface e.g., the best setting for the robot setup is an initial electron energy of 280 keV and 17 cm target distance. This setting provides the largest uniform dose distribution and the sharpest energy distribution with a FWHM of 36 keV at the sample surface. Figure 7 shows the ECSS beam uniformity area and its relative deviations from the mean value.

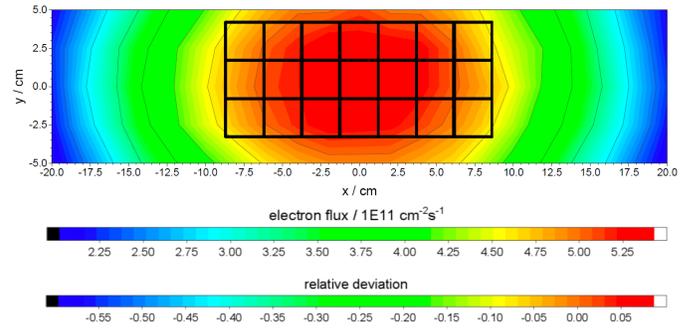

Fig.7 Example of an electron flux distribution of the static robot setup with indicated sample area inside the irradiation box, absolute electron flux values and relative deviations. The FWHM of energy distribution on the target surface is 36 keV and the secondary photon contribution is less than 0.1 %.

## IV. EXPERIMENTAL PROCEDURE

Prior to each irradiation session, a dosimetry run with radiochromic film is taken. The dosimetric data is then used for fine adjustment of the beam current to correct for small fluctuations of the beam generator. Target values for the dosimetry can be obtained using a conversion table for the dose rate, total dose, particle flux, duration and fluence taking into account the energy and material-dependent mass-stopping power of accelerated electrons [7]. An example is given in Table 2.

| Electron Energy (keV) | dE/dX · 1/ρ (col. st. pw.) [MeV/(g/cm²)] for PMMA | (Incident) Flux on absorber [$N_e$/(cm²·s)] | Fluence on absorber [$N_e$/cm²] | Duration [mins] | Delivered Total Dose [Gy] | Dose Rate [Gy/min] | Mass per Area 3 mm PMMA [g/cm²] | Mass per Area of e-exit foil [g/cm²] | CSDA range [g/cm²] |
|---|---|---|---|---|---|---|---|---|---|
| 1000 | 1,788 | 7,28E+11 | 3,15E+17 | 7200,09 | 9,00E+07 | 1,25E+04 | 4,26E-01 | 2,26E-02 | 4,50E-01 |
| 1000 | 1,788 | 8,74E+11 | 3,15E+17 | 6000,76 | 9,00E+07 | 1,50E+04 | 4,26E-01 | 2,26E-02 | 4,50E-01 |

Table 2. Conversion table used to determine the flux, fluence and run duration.

After determining the adjusted beam parameters, the irradiation box can be equipped with the samples and closed. If the treatment shall be done under an inert atmosphere, nitrogen purging can be activated. The online monitor supervises the elimination from residual oxygen from the box. After establishing the desired conditions, the irradiation may start. During the warm-up procedure of the electron generators, the irradiation box can be placed at a distance (or the robot arm with source is moved away from the box) in a way that the samples see no irradiation. Another option is to shield the box from the electron flux during this period. After the warm-up the samples are irradiated for the defined duration with a second resolution. All irradiation conditions are monitored in real-time by the staff of the facility and are available for retrospective analysis and evaluation.

## V. CONCLUSION

The collaboration has made it possible to adapt the existing irradiation setups, mainly used for applications focusing on the study and modification of polymer materials, into a facility where electrons of high intensity and low energy can be applied on large surfaces following relevant ECSS/ESA requirements.

These sites are currently being used by major aerospace players on a regular basis. Many different typology samples have already been tested for large European space exploration programs. The experimental setups provide stable electron beams with minimal energy straggling, a large irradiation area with <10% non-uniformity, excellent thermal control of samples and inert gas atmosphere with controlled oxygen contamination. In addition, these sites may also be used for total ionizing dose tests with electrons according to ECSS-22900.

REFERENCES AND FOOTNOTES

A. *References*

B. *Footnotes*